\documentclass[runningheads]{llncs}

\usepackage{amsfonts}
\usepackage{epsfig}
\usepackage{amsmath}
\usepackage{graphicx}
\usepackage{amsmath}
\usepackage{amscd}
\usepackage{stmaryrd}
\usepackage{algorithm}
\usepackage{arydshln}
\usepackage{algorithmic}
\usepackage{youngtab}
\usepackage{young}
\usepackage{geometry}
\usepackage{bbm}\usepackage{multirow}
\usepackage{leftidx}\usepackage{mathdots}
\usepackage{bm}
\usepackage{extarrows}
\usepackage{booktabs}

\usepackage[colorlinks,
            linkcolor=red,
            anchorcolor=red,
            citecolor=red,
            ]{hyperref}

\def\bt{\begin{theorem}}
\def\et{\end{theorem}}
\def\bp{\begin{proposition}}
\def\ep{\end{proposition}}
\def\bc{\begin{corollary}}
\def\ec{\end{corollary}}
\def\bo{\begin{proof}}
\def\eo{\end{proof}}
\def\bx{\begin{example}}
\def\ex{\end{example}}
\def\br{\begin{remark}}
\def\er{\end{remark}}
\def\bl{\begin{lemma}}
\def\el{\end{lemma}}

\def\bn{\begin{definition}}
\def\en{\end{definition}}
\def\ba{\begin{array}}
\def\ea{\end{array}}
\def\be{\begin{equation}}
\def\ee{\end{equation}}
\def\bd{\begin{description}}
\def\ed{\end{description}}
\def\bu{\begin{enumerate}}
\def\eu{\end{enumerate}}
\def\bi{\begin{itemize}}
\def\ei{\end{itemize}}

\newbox\bigstrutbox
\setbox\bigstrutbox=\hbox{\vrule height12.5pt depth5pt width0pt}
\def\bigstrut{\relax\ifmmode\copy\bigstrutbox\else\unhcopy\bigstrutbox\fi}

\newbox\Bigstrutbox
\setbox\Bigstrutbox=\hbox{\vrule height17.5pt depth5pt width0pt}
\def\Bigstrut{\relax\ifmmode\copy\Bigstrutbox\else\unhcopy\Bigstrutbox\fi}

\def\i{{\bf i}}

\def\x{{\bf x}}
\def\y{{\bf y}}
\def\z{{\bf z}}

\def\0{{\bf 0}}
\def\1{{\bf 1}}
\def\2{{\bf 2}}
\def\3{{\bf 3}}
\def\4{{\bf 4}}
\def\5{{\bf 5}}
\def\6{{\bf 6}}
\def\7{{\bf 7}}
\def\8{{\bf 8}}
\def\9{{\bf 9}}

\def\ds{\displaystyle}

\begin{document}

\pagestyle{headings}

\mainmatter

\title{Quantum Algorithms to Matrix Multiplication}

\author{Changpeng Shao \\
cpshao@amss.ac.cn}

\authorrunning{Changpeng Shao}

\institute{Academy of Mathematics and Systems Science, Chinese Academy of Sciences, Beijing 100190, China}

\maketitle

\begin{abstract}
In this paper, we study quantum algorithms of matrix multiplication from the viewpoint of inputting quantum/classical data to outputting quantum/classical data. The main target is trying to overcome the input and output problem, which are not easy to solve and many quantum algorithms will encounter, to study matrix operations in quantum computer with high efficiency. And solving matrix multiplication will be the first step. We propose three quantum algorithms to matrix multiplication based on swap test, SVE and HHL. From the point of making fewer assumptions, swap test method works the best than the other two. We also show that the quantum algorithm of matrix multiplication with classical input and output data by swap test achieves the best complexity $\widetilde{O}(n^2/\epsilon)$ with no assumptions. This is proved by giving an efficient quantum algorithm in polynomial time to solve the input problem, that is to prepare the quantum states of the classical data efficiently. Other contributions of this paper include:
(1). Extending swap test to a more general form that is suitable to deal with quantum data in parallel, which will have further applications in other matrix operations.
(2). Generalizing SVE technique such that it applies to any matrix (not just Hermitian) directly only with quantum data.
(3). Proposing two new efficient quantum algorithms to prepare quantum states of classical data, which solves the input problem efficiently than other quantum algorithms.
\end{abstract}


\section{Introduction}
\setcounter{equation}{0}

In the study of quantum algorithms (for example, see \cite{biamonte}, \cite{harrow}, \cite{wiebe}), people usually encounter the ``input" and ``output" problem. The input problem is the transformation from classical data (such as complex vectors) into quantum data (such as quantum states); the output problem is the converse. These two problems generally are not easy to solve efficiently in quantum computer, and sometimes even cost more than other steps of the quantum algorithms. So mostly we just assume that we already get the quantum data by some methods when studying quantum algorithms. Another important idea is studying quantum algorithms with input and output data are all quantum, such as the quantum machine learning for quantum data  \cite{biamonte}, principal component analysis \cite{lloyd-pca}, quantum simulator \cite{marvian}, \cite{wiebe-simulator} and so on.

In this paper, we study the matrix multiplication in quantum computer from two different perspectives with three different techniques. The two perspectives share the same input data, that is it can be classical or quantum. The difference reflects in the output, one is quantum and the other one is classical. The three techniques we will use are swap test \cite{buhrman}, SVE \cite{kerenidis} and HHL \cite{harrow}. In the two different perspectives, if the input data is quantum, then we can only apply swap test and SVE; if the input data is classical, then all the three techniques can play roles in. What we cares more in this work is quantum data to quantum data. Although SVE contains more wide applications, its performance in matrix multiplication is not efficient than swap test. Also to make swap test works for matrix multiplication from quantum data to quantum data, a more generalized version of swap test (proposition \ref{prop-swap test}) will be proposed in this paper. Note that the swap test proposed in \cite{buhrman} can be viewed as a procedure from quantum data to classical data, now the new version will achieve quantum data to quantum data.

The target of this study aims at extending classical matrix operations into quantum case (i.e., quantum input and output), hoping to obtain efficient matrix operations in quantum computer and so solving the classical problems more efficiently. Also for the comparison with classical algorithms to matrix multiplication, quantum algorithm to matrix multiplication achieve classical data to classical data are also studied comprehensively in this work, which contains two sub-works: the preparation of quantum states (classical data to quantum data) and the reading out from quantum data into classical data.

As for the matrix multiplication, the reading out problem is not difficult to solve mainly based on swap test. However, the preparation of quantum states is not so easy in quantum computer generally. Efficient quantum algorithms to certain special cases still exists (for instance, see \cite{clader}, \cite{grover-state-preparation}, \cite{lloyd}, \cite{soklakov}). One special well known case is when the classical data is relatively uniform distributed \cite{aaronson}, \cite{clader}, \cite{lloyd}. Based on this special case, in the paper we will propose two new quantum algorithms (theorem \ref{state prepare: indirect method 2} and \ref{state prepare: indirect method 3}) to prepare the quantum states of classical data in the general case. The corresponding complexities are satisfactory and better than any other quantum algorithm to achieve the same target to my knowledge. When we are given enough information (such as the maximum, the minimum, the norm, the positions of nonzero entries and so on) about the classical data, then the input problem can be solved efficiently in polynomial time. Obtaining such information may take some extra time. In the matrix multiplication problem, however, all these required information can be obtained before implementing the quantum algorithms. If the given matrix is $n$-by-$n$, then to get the required information will take at most $O(n^2)$, which is acceptable, since $\Omega(n^2)$ is the lower bound of matrix multiplication problem. This means, the input problem can be actually solved ``efficiently" in matrix multiplication.

The obtained quantum algorithms to matrix multiplication is polynomially depends on the precision. Therefore, when the precision is bounded by $O(1/\textmd{poly}\log n)$, then quantum computer can solve the matrix multiplication problem efficiently in time $\widetilde{O}(n^2)$ by swap test (see table \ref{classical-data-simplified}). The quantum algorithms obtained from SVE and HHL depend on the condition number of the given matrices. If the condition number is bounded by $O(\textmd{poly}\log n)$, then these two algorithms also achieve the best efficiency. However, if we only interested in quantum output data, then all the three quantum algorithms are rely on the condition number. With the same assumption above on precision and condition number, if the input is classical data, then this problem can be solved in polynomial time by HHL; if the input is quantum data, then it can be solved in time $\widetilde{O}(\sqrt{n})$ (see table \ref{quantum-data}) by swap test and SVE.

The structure of this paper is as follows: In section 2, 3, we consider the quantum algorithms to the matrix multiplication with quantum output and classical output under one assumption about the preparation of quantum states. Section 4 devotes to study efficient quantum algorithms to solve the assumption.

{\bf Notations.}
In this paper, $\i$ refers to the imaginary unit $\sqrt{-1}$. For any $n\times n$ matrix $A=(a_{ij})$, the notation $\|A\|_F:=\sqrt{\sum_{i,j}|a_{ij}|^2}$ refers to the Frobenius norm of $A$; for any vector $\x=(x_0,\ldots,x_{n-1})$, the notation $\|\x\|_p:=(\sum_i |x_i|^p)^{1/p}$ refers to the $p$ norm of $\x$.

\section{Quantum algorithm to matrix multiplication with quantum information}
\setcounter{equation}{0}

This section is devoted to study matrix multiplication with quantum output data. The problem can be stated as: Given two matrices $A,B$ or their quantum states by viewing them as vectors (see \eqref{ass1} below), we want to get the quantum state of $AB$. First, we make some statements of the preliminaries.

Let $A=(a_{ij})_{n\times n}$ be a given $n$-by-$n$ matrix. We denote its $i$-th row as $A_{i\bullet}$, the $j$-th column as $A_{\bullet j}$.
In order to study matrix multiplication in quantum computer, we should have some quantum information about given classical matrices.
So we make the following assumption in this and the next section:

\begin{description}
  \item[Assumption:] Assume that we have efficient quantum algorithm to achieve the quantum state preparation of the rows and the columns of $A$, such as by QRAM \cite{giovannetti} or the data structure introduced in \cite{kerenidis}. That is we can get
  \be \label{ass1}
  |A\rangle=\frac{1}{\|A\|_F}\sum_{i,j=0}^{n-1}a_{ij}|i,j\rangle
  =\frac{1}{\|A\|_F}\sum_{i=0}^{n-1}\|A_{i\bullet}\|_2|i\rangle|A_{i\bullet}\rangle
  =\frac{1}{\|A\|_F}\sum_{j=0}^{n-1}\|A_{\bullet j}\|_2|A_{\bullet j}\rangle|j\rangle
  \ee
  efficiently, where $\|A\|_F=\sqrt{\sum_{i,j}|a_{ij}|^2}$ refers to the Frobenius norm of $A$.
\end{description}

\br \label{remark-input-data}
By applying the inverse of the quantum algorithm of preparing the quantum states of  rows and columns of $A$ on $|A\rangle$, we can obtain the following quantum states
\be \label{ass2}
|A_{F\bullet} \rangle=\frac{1}{\|A\|_F}\sum_{i=0}^{n-1}\|A_{i\bullet}\|_2 |i\rangle, \hspace{.3cm}
|A_{\bullet F}\rangle=\frac{1}{\|A\|_F}\sum_{j=0}^{n-1}\|A_{\bullet j}\|_2 |j\rangle.
\ee

Also note that, the density matrix of $|A\rangle$ equals
$
\frac{1}{\|A\|_F^2}\sum_{i=0}^{n-1}\|A_{i\bullet}\|_2^2|i\rangle|A_{i\bullet}\rangle\langle i|\langle A_{i\bullet}|.
$
By taking the trace on the second register, we will get the density matrix
of $|A_{F\bullet} \rangle$, which is another method to get the quantum state of (\ref{ass2}) from (\ref{ass1}).
For convenience, we define $A_{F\bullet}$ and $A_{\bullet F}$ as the column vectors with entries $\|A_{i\bullet}\|_2$ and $\|A_{\bullet j}\|_2$ respectively.
\er

\br
The preparation of quantum states work efficient in practical in some cases \cite{clader}, \cite{grover-state-preparation}, \cite{lloyd}, \cite{soklakov}. We will discuss this in section \ref{Preparation of quantum states}. Two new efficient quantum algorithms to prepare quantum state will be proposed then, which makes the assumption above more reliable.
\er

Given two matrices $A,B$, in the following, we first consider the quantum algorithm to prepare the quantum state of $AB$ from swap test by giving the quantum information of $A$ and $B$. Certainly, this problem can be solved by the singular value estimation technique (SVE) proposed in \cite{kerenidis} still with quantum information of $A$ and $B$. Moreover, if we view $A$ as a classical data, then the quantum state of $AB$ can also be solved by performing the matrix multiplication algorithm obtained from HHL algorithm \cite{harrow}. The complexity of the algorithms obtained in the last two ways depends on the condition number of $A$ or $B$. Also in HHL algorithm, we need the Hamiltonian simulation of
$\widetilde{A}=\left(
   \begin{array}{cc}
     0 & A \\
     A^\dagger & 0 \\
   \end{array}
 \right)$ to be efficient.

We only consider the case of real matrix multiplication, as for complex matrix multiplication, we just need to focus on the real and imaginary parts separately.

\subsection{By swap test}

{\bf 2.1.1 \hspace{0.14cm} Swap test and its generalization} \vspace{.2cm}
\\
Swap test  was first proposed in \cite{buhrman} as a byproduct of quantum phase estimation algorithm and Grover searching, which can be used to compute the probability of some desired quantum states. It plays an important role in many quantum algorithma, such as in HHL algorithm \cite{harrow} and some machine learning algorithms \cite{lloyd}, \cite{rebentros}, \cite{schuld}, \cite{wang}, \cite{wiebe}, to estimate the inner product of two quantum states.
In the following, we first briefly review the underlying problem swap test considers and the basic procedures to solve it.
Notice that swap test can only returns the classical information (i.e., the inner product of two quantum states we want) if we perform a measurement,
which is not enough to deal with quantum data in matrix multiplication in parallel, even if we do not perform measuring.
To overcome this problem, we extend swap test into a more general form (proposition \ref{prop-swap test} below) that can output quantum data in parallel, which may has other applications except matrix multiplication considered in this paper.

Let
\be\label{ini-form}
|\phi\rangle=\sin\theta|0\rangle|u\rangle+\cos\theta|1\rangle|v\rangle
\ee
be a unknown quantum state that can be prepared in time $O(T_{\textmd{in}})$, where $|u\rangle,|v\rangle$ are normalized quantum states. We want to consider the problem that how to estimate $\theta$ in quantum computer to accuracy $\epsilon$ with high success probability at least $1-\delta$.

Suppose that $|\phi\rangle$ comes from some algorithms, that means there is a given unitary $U$ such that $|\phi\rangle=U|0\rangle$.
Let $Z$ be the 2-dimensional unitary transformation that maps $|0\rangle$ to $-|0\rangle$ and $|1\rangle$ to $|1\rangle$, which is usually called Pauli-Z matrix. Denote $G=(2|\phi\rangle\langle\phi|-I)(Z\otimes I)$, which is the rotation matrix used in Grover's searching algorithm. Then
\[
G=\left(
   \begin{array}{rr} \vspace{.2cm}
     \cos2\theta  &~~ \sin2\theta \\
     -\sin2\theta &~~ \cos2\theta \\
   \end{array}
 \right)
\]
in the basis $\{|0\rangle|u\rangle,|1\rangle|v\rangle\}$. The eigenvalues of $G$ are $e^{\pm\i2\theta}$ and the corresponding eigenvectors are
\[
|w_1\rangle=\frac{1}{\sqrt{2}}\Big(|0\rangle|u\rangle+\i|1\rangle|v\rangle\Big),~~|w_2\rangle=\frac{1}{\sqrt{2}}\Big(|0\rangle|u\rangle-\i|1\rangle|v\rangle\Big)
\]
respectively. Note that
$
|\phi\rangle=-\frac{\i}{\sqrt{2}} (e^{\i\theta}|w_1\rangle-e^{-\i\theta}|w_2\rangle ).
$
So performing quantum phase estimation algorithm on $G$ with initial state $|0\rangle^n|\phi\rangle$ for some $n=O(\log1/\delta\epsilon)$.
We will get an approximate of the following state
\be\label{final-form}
-\frac{\i}{\sqrt{2}}\Big(e^{\i\theta}|y\rangle|w_1\rangle-e^{-\i\theta}|-y\rangle|w_2\rangle\Big),
\ee
where $y\in \mathbb{Z}_{2^n}$ satisfies $|\theta-y\pi/2^n|\leq \epsilon$. The time complexity of the above procedure is $O(T_{\textmd{in}}/\epsilon\delta)$. Sometimes, $\delta$ will be ignored just for simplicity in the complexity analysis. Performing a measurement on (\ref{final-form}), we will get an $\epsilon$ approximate of $\theta$.

Furthermore, let $f(y)=g(\theta)$ be some functions such that $f(y)=f(-y)$ (i.e., $f$ is an even function), then from (\ref{final-form}), we can get
\be\label{final-form-1}
|g(\theta)\rangle|\phi\rangle,
\ee
by adding a register to store $g(\theta)$ and undoing the quantum phase estimation. This is a procedure that we want to further make use of more quantum information about $\theta$ instead of outputting.

Now let $|x\rangle,|y\rangle$ be two real quantum states, except a global phase, which can be prepared in time $O(T_{\textmd{in}})$. Then the above method provide us an quantum algorithm to estimate $\langle x|y\rangle$ to accuracy $\epsilon$ in time $O(T_{\textmd{in}}/\epsilon)$. Actually, we just need to consider the state
\[
|\phi\rangle=\frac{1}{\sqrt{2}}(|+\rangle|x\rangle+|-\rangle|y\rangle)
=\frac{1}{2}(|0\rangle(|x\rangle+|y\rangle)+|1\rangle(|x\rangle+|y\rangle)).
\]
The probability of $|0\rangle$ (resp. $|1\rangle$) is $(1+\langle x|y\rangle)/2$ (resp. $(1-\langle x|y\rangle)/2$). So we can set $\sin\theta=\sqrt{(1+\langle x|y\rangle)/2}$ and $\cos\theta=\sqrt{(1-\langle x|y\rangle)/2}$. The quantum state $|\phi\rangle$ can be rewritten in the form (\ref{ini-form}), where $|u\rangle,|v\rangle$ corresponds to the normalization of $|x\rangle+|y\rangle,|x\rangle-|y\rangle$. Therefore, the inner product $\langle x|y\rangle$ can be evaluated in time $O(T_{\textmd{in}}/\epsilon)$ with accuracy $\epsilon$. Concluding this, we get the following result

\bp \label{cor:inner product}
Let $|x\rangle,|y\rangle$ be two quantum states, which can be prepared in time $O(T_{\emph{in}})$,
then $\langle x|y\rangle$ can be estimated with accuracy $\epsilon$ in time $O(T_{\emph{in}}/\epsilon)$.
\ep

\br
If $|x\rangle,|y\rangle$ are complex quantum states, then the probability of $|0\rangle$ (resp. $|1\rangle$) is
$(1+\textmd{Re}\langle x|y\rangle)/2$ (resp. $(1-\textmd{Re}\langle x|y\rangle)/2$). So, we can only get the value of $\textmd{Re}\langle x|y\rangle$.
The image part of $\langle x|y\rangle$ can be computed by considering the inner product of $|x\rangle$ with $\i|y\rangle$.
\er

The above method to estimate $\langle x|y\rangle$  is usually called swap test \cite{buhrman}. Note that quantum counting \cite{brassard} can also used to estimate $\langle x|y\rangle$. They contain the same idea. Moreover, from (\ref{final-form-1}), we actually can obtain the following quantum state
\[
\frac{1}{\sqrt{2}}\Big(|0\rangle|x\rangle+|1\rangle|y\rangle\Big)\Big|g(\langle x|y\rangle)\Big\rangle,
\]
for any function $g$, since cosine function is even.

\bp \label{prop-swap test}
Let $|x\rangle,|y\rangle$ be two real quantum states, except a global phase, which can be prepared in time $O(T_{\emph{in}})$. Let $f$ be any function. Then there is a quantum algorithm within time $O(T_{\emph{in}}/\epsilon)$ to achieve
\be \label{swap test:general procedure}
\frac{1}{\sqrt{2}}(|0\rangle|x\rangle+|1\rangle|y\rangle)\mapsto
\frac{1}{\sqrt{2}}(|0\rangle|x\rangle+|1\rangle|y\rangle)|f(s)\rangle,
\ee
where $|\langle x|y\rangle-s|\leq \epsilon$.
\ep

From proposition \ref{prop-swap test}, it is easy to get the following result

\bc
For any given quantum state $\sum_j\alpha_j|j\rangle$ prepared in time $O(T_{\emph{in}})$ and any function $f$, we can obtain $\sum_j\alpha_j|j\rangle|f(\tilde{\alpha}_j)\rangle$ in time $O(T_{\emph{in}}/\epsilon),$ where $|\alpha_j-\tilde{\alpha}_j|\leq \epsilon$.
\ec

For instance, in HHL algorithm to solve the linear system $Ax=b$. When we get $|x\rangle=\sum x_i|i\rangle$, then by the above corollary, we can change it into $\sum x_i|i\rangle|f(x_i)\rangle$. So we can further apply the quantum information of $x_i$ concurrently for other problems.
\vspace{.4cm}
\\
{\bf 2.1.2  \hspace{0.14cm} Matrix multiplication algorithm by swap test} \vspace{.2cm}
\\
With the above preliminaries about swap test, now we can consider the matrix multiplication problem. The basic idea is similar to procedure \eqref{swap test:general procedure} by putting the inner product of quantum states into another register, then like the procedure of HHL algorithm to put this value into the coefficient. At this time, the quantum parallelism will play an important role in helping us deal with the inner product in parallel.

Denote $C=AB$, the target of the following quantum algorithm aim at finding the quantum information of $C$, that is $|C\rangle$. Returning classical information of matrix multiplication will be studied in the next section.
Note that $C_{ij}=A_{i\bullet}^T B_{\bullet j}=\|A_{i\bullet}\|_2\|B_{\bullet j}\|_2 \langle A_{i\bullet}|B_{\bullet j}\rangle$. By swap test introduced above, we can estimate $\langle A_{i\bullet}|B_{\bullet j}\rangle$ efficiently. Together with quantum parallelism, we can get the desired quantum state $|C\rangle$ efficiently in the following five steps:

Step 1, consider the initial state, which equals the tensor product of $|A_{F\bullet} \rangle$ and $|B_{\bullet F}\rangle$:
\[
\ds\frac{1}{\|A\|_F\|B\|_F}\sum_{i,j=0}^{n-1}\|A_{i\bullet}\|_2\|B_{\bullet j}\|_2 |i,j\rangle|0,0\rangle.
\]

Step 2, by control transformation, we can prepare $|A_{i\bullet}\rangle$ and $|B_{\bullet j}\rangle$ in the last register:
\be\ba{lll}\label{state2} \vspace{.2cm}
&&\ds\frac{1}{\|A\|_F\|B\|_F}\sum_{i,j=0}^{n-1}\|A_{i\bullet}\|_2\|B_{\bullet j}\|_2 |i,j\rangle
\otimes \frac{1}{\sqrt{2}}\Big(|0\rangle|A_{i\bullet}\rangle+|1\rangle|B_{\bullet j}\rangle\Big) \\
&\mapsto&\ds\frac{1}{\|A\|_F\|B\|_F}\sum_{i,j=0}^{n-1}\|A_{i\bullet}\|_2\|B_{\bullet j}\|_2 |i,j\rangle
\otimes \frac{1}{2}\Big(|0\rangle(|A_{i\bullet}\rangle+|B_{\bullet j}\rangle)+|1\rangle(|A_{i\bullet}\rangle-|B_{\bullet j}\rangle)\Big).
\ea\ee

Denote $|\phi_{ij}\rangle=\frac{1}{2}(|0\rangle(|A_{i\bullet}\rangle+|B_{\bullet j}\rangle)+|1\rangle(|A_{i\bullet}\rangle-|B_{\bullet j}\rangle))
=\sin\theta_{ij}|0\rangle|u_{ij}\rangle+\cos\theta_{ij}|1\rangle|v_{ij}\rangle$, where $\sin^2\theta_{ij}$ (resp. $\cos^2\theta_{ij}$) is the probability of $|0\rangle$ (resp. $|1\rangle$) and $|u_{ij}\rangle$  (resp. $|v_{ij}\rangle$) is the normalization of $|A_{i\bullet}\rangle+|B_{\bullet j}\rangle$ (resp. $|A_{i\bullet}\rangle-|B_{\bullet j}\rangle$).
Also denote the eigenvalues of $G_{ij}=(2|\phi_{ij}\rangle\langle\phi_{ij}|-I)(Z\otimes I)$ as $e^{\pm\i2\theta_{ij}}$ and the corresponding eigenvectors as $|w_{ij}^\pm\rangle$. Then (\ref{state2}) can be written as
\[
\frac{1}{\|A\|_F\|B\|_F}\sum_{i,j=0}^{n-1}\|A_{i\bullet}\|_2\|B_{\bullet j}\|_2 |i,j\rangle
\Big(\sin\theta_{ij}|0\rangle|u_{ij}\rangle+\cos\theta_{ij}|1\rangle|v_{ij}\rangle\Big).
\]

Step 3, perform quantum phase estimation to $G_{ij}$ with the initial state $(\sin\theta_{ij}|0\rangle|u_{ij}\rangle+\cos\theta_{ij}|1\rangle|v_{ij}\rangle)|0\rangle$. Together with the control operation, we can get
\[
\ds\frac{-\i}{\sqrt{2}\|A\|_F\|B\|_F}\sum_{i,j=0}^{n-1}\|A_{i\bullet}\|_2\|B_{\bullet j}\|_2 |i,j\rangle
\Big(e^{\i\theta_{ij}}|w_{ij}^+\rangle|y_{ij}\rangle-e^{-\i\theta_{ij}}|w_{ij}^-\rangle|-y_{ij}\rangle\Big),
\]
where $y_{ij}\pi/2^n$ is a good approximate of $\theta_{ij}$ to accuracy $\epsilon$.

Step 4, apply control rotation based on the register $|\pm y_{ij}\rangle$, which returns the following state
\[\ba{lll}\vspace{.2cm}
&& \ds\frac{-\i}{\sqrt{2}\|A\|_F\|B\|_F}\sum_{i,j=0}^{n-1}\|A_{i\bullet}\|_2\|B_{\bullet j}\|_2 |i,j\rangle
\Big(e^{\i\theta_{ij}}|w_{ij}^+\rangle|y_{ij}\rangle-e^{-\i\theta_{ij}}|w_{ij}^-\rangle|-y_{ij}\rangle\Big) \hspace{1cm}  \\
&& \hfill \otimes \Big(\langle A_{i\bullet}|B_{\bullet j}\rangle|0\rangle+\sqrt{1-\langle A_{i\bullet}|B_{\bullet j}\rangle^2}|1\rangle\Big).
\ea\]

Step 5, undo the procedure 1-3, which yields the desired state
\[
\frac{1}{\|A\|_F\|B\|_F}\sum_{i,j=0}^{n-1}\|A_{i\bullet}\|_2\|B_{\bullet j}\|_2 \langle A_{i\bullet}|B_{\bullet j}\rangle |i,j\rangle|0\rangle+|0\rangle^\bot.
\]

The next thing we need to do is estimating the error and the final complexity. This procedure is quite simple just based on triangle inequality of norm, so we put all the details in appendix \ref{app1}.  Note that the above algorithm procedure hold for all matrices, not just square. The final result can be summarized in the following

\bt \label{result1}
For any two matrices $A, B$,  the quantum state of $AB$ can be obtained in time
$
\widetilde{O}(\|A\|_F^3\|B\|_F^3/\|AB\|_F^3\epsilon)
$
to accuracy $\epsilon$.
\et

For instance,

(1). If $B=|b\rangle$, then the quantum state $|Ab\rangle$ can be obtained in time
$\widetilde{O}(\|A\|_F^2/\epsilon\|A|b\rangle\|_2^2)=\widetilde{O}(n\kappa^2/\epsilon)$, where $\kappa$ is the condition number of $A$. Actually, the result in theorem \ref{result1}  is also bounded by $\widetilde{O}(n\kappa^2/\epsilon)$.

(2). If $A=|a\rangle$ and $B=\langle b|$, then the quantum state of the the rank 1 matrix $|a\rangle\langle b|$ can be obtained in time $\widetilde{O}(1/\epsilon)$. This result also holds when $A,B$ are given in classical column and row vectors.

(3). For any two general matrices $A,B$, they can decomposed by columns and rows, that is $A=(A_{\bullet 0},\ldots,A_{\bullet (n-1)})$ and $B=(B_{0\bullet}^T,\ldots,B_{(n-1)\bullet}^T)^T$. Then $AB=\sum_j A_{\bullet j}B_{j \bullet}^T$. Since the quantum state of $|A_{\bullet j}\rangle\langle B_{j \bullet}|$ to accuracy $\epsilon$, denoted as $|\widetilde{C}_j\rangle$, can be obtained in time $\widetilde{O}(1/\epsilon)$. Then we just need to compute the quantum state $|\widetilde{C}\rangle$ proportional to the linear combination
$\sum_j\|A_{\bullet j}\|_2\|B_{j\bullet}\|_2|\widetilde{C}_j\rangle$. In \cite[Chapter 26]{childs}, there is a quantum algorithm to get $|\widetilde{C}\rangle$ in time
\[
\widetilde{O}\left( \frac{\sum_j\|A_{\bullet j}\|_2\|B_{j\bullet}\|_2 }{\epsilon\left\|\sum_j\|A_{\bullet j}\|_2\|B_{j\bullet}\|_2|\widetilde{C}_j\rangle\right\|_2} \right)
=\widetilde{O}\left( \frac{ A_{\bullet F}\cdot B_{F\bullet} }{\epsilon\|AB\|_F} \right),
\]
where $A_{\bullet F}=(\|A_{\bullet 0}\|_2,\ldots,\|A_{\bullet (n-1)}\|_2)^T$ and $B_{F\bullet}=(\|B_{0\bullet}\|_2,\ldots,\|B_{(n-1)\bullet}\|_2)^T$
(see remark \ref{remark-input-data}).
The above result can also changed only depending the Frobenius norm by Cauchy inequality  into  the form
$\widetilde{O}(\|A\|_F\|B\|_F/\epsilon \|AB\|_F)$, which is better than theorem \ref{result1}.

\bt \label{result1:better}
Let $A$ be an $l\times m$ matrix and $B$ an $m\times n$ matrix such that $ln\neq 1$, then the quantum state of $AB$ can be obtained in time
$
\widetilde{O}(\|A\|_F\|B\|_F/\|AB\|_F\epsilon)
$
to accuracy $\epsilon$.
\et

\subsection{By SVE and HHL}

In \cite{kerenidis}, Kerenidis et al introduced a data structure, which is similar to QRAM \cite{giovannetti}, to store classical matrices in quantum computer efficiently. Based on this data structure, a fast quantum algorithm to the singular value estimated (SVE for brief, which is close to singular value decomposition) was obtained. More precisely, if $A=\sum\sigma_i|u_i\rangle\langle v_i|$ is the singular value decomposition of $A$, then there is an efficient quantum algorithm to achieve $\sum \alpha_i|v_i\rangle\mapsto \sum \alpha_i|v_i\rangle|\sigma_i\rangle$. This algorithm is enough to solve certain problems relating to matrix operations, like multiplication or inversion. Moreover, their algorithm about SVE only applies the quantum information of $A$.

In this subsection, we first review the basic ideas of SVE, then we generalize their result into a quantum algorithm to achieve
$\sum \alpha_i|v_i\rangle\mapsto \sum \alpha_i|u_i\rangle|\sigma_i\rangle$. So even when $A$ is not Hermitian, we can also perform matrix multiplication or inversion directly only with the quantum information of $A$.
Although, the data structure proposed in \cite{kerenidis} lie in a model which is a little different from the standard quantum circuit model, their result about SVE only depends on the efficient preparation of some quantum states. So with the assumption given  in the beginning of this section, their result also works in the standard quantum circuit model.

Let $A=(a_{ij})_{n\times n}$ be a $n\times n$ matrix, based on the assumption given in the beginning of this section, we know that
quantum computer can perform the following mappings efficiently in time $O({\rm poly}\log(n))$:
\[
U_\mathcal{M}:|i\rangle|0\rangle\mapsto|i\rangle|A_{i\bullet}\rangle=\ds\frac{1}{\|A_{i\bullet}\|}\sum_{j=0}^{n-1}a_{ij}|i,j\rangle, \hspace{1cm}
U_\mathcal{N}:|0\rangle|j\rangle\mapsto|A_{F\bullet}\rangle|j\rangle=\ds\frac{1}{\|A\|_F}\sum_{j=0}^{n-1}\|A_{i\bullet}\||i,j\rangle.
\]

\br
The mapping $U_\mathcal{M}$ and $U_\mathcal{N}$ seem too perfect. Generally, the results will contain some other orthogonal parts or some errors in the results. However, by amplitude amplification, we can make it very close to the results given in the above formula. In the following, to make things simple, we just use these two mappings as \cite{kerenidis} did.
\er

Define two degenerate operators $\mathcal{M}$ and $\mathcal{N}$ as:
$\mathcal{M}:|i\rangle\mapsto|i\rangle|A_{i\bullet}\rangle,$ and
$\mathcal{N}:|j\rangle\mapsto|A_{F\bullet}\rangle|j\rangle.$
Then $\mathcal{M}^\dagger\mathcal{N}=\frac{A}{\|A\|_F}.$
It is also easy to check that $\mathcal{M}^\dagger\mathcal{M}=\mathcal{N}^\dagger\mathcal{N}=I_n$. The reflections
$2\mathcal{M}\mathcal{M}^\dagger-I_{n^2}$ and $2\mathcal{N}\mathcal{N}^\dagger-I_{n^2}$ can be efficiently implemented in quantum computer. Denote $W=(2\mathcal{M}\mathcal{M}^\dagger-I_{n^2})(2\mathcal{N}\mathcal{N}^\dagger-I_{n^2})$.
Let $A=\sum_{i=0}^{n-1} \sigma_i|u_i\rangle\langle v_i|$ be the singular value decomposition of $A$, then
\[\ba{lll}\vspace{.2cm}
W\mathcal{N}|v_i\rangle &=& \ds\frac{2\sigma_i}{\|A\|_F}\mathcal{M}|u_i\rangle-\mathcal{N}|v_i\rangle; \\
W\mathcal{M}|u_i\rangle &=& \ds\Big(\frac{4\sigma_i^2}{\|A\|_F^2}-1\Big)\mathcal{M}|u_i\rangle-\frac{2\sigma_i}{\|A\|_F}\mathcal{N}|v_i\rangle.
\ea\]
So the subspace span$\{\mathcal{M}|u_i\rangle,\mathcal{N}|v_i\rangle\}$ is invariant under $W$. The matrix representation of $W$ in this space is
\[
W_i=\left(
      \begin{array}{cc}\vspace{.3cm}
        \ds\frac{4\sigma_i^2}{\|A\|_F^2}-1 &~~ \ds\frac{2\sigma_i}{\|A\|_F} \\
       \ds -\frac{2\sigma_i}{\|A\|_F}      &~~ -1 \\
      \end{array}
    \right).
\]
The eigenvalues of $W_i$ are
$\exp(\pm \i\theta_i)$
where $\theta_i$ satisfies
$
\cos\theta_i= {2\sigma_i^2}/{\|A\|_F^2}-1.
$
So $\cos(\theta_i/2)={\sigma_i}/{\|A\|_F}$. The corresponding eigenvectors are
$
w^{\pm}_i=-\mathcal{M}|u_i\rangle+e^{\mp \i \theta_i/2}\mathcal{N}|v_i\rangle.
$
It is easy to get the following decomposition
\[
\mathcal{N}|v_i\rangle=\frac{1}{2\i \sin(\theta_i/2)} (w^+-w^-), \hspace{.5cm}
\mathcal{M}|u_i\rangle=\frac{1}{2\i \sin(\theta_i/2)} (e^{\i\theta_i/2}w^+-e^{-\i\theta_i/2}w^-).
\]
With the above notations, we now can prove a more general result than \cite{kerenidis}.

\bp \label{svd}
Let $A$ be a $n\times n$ matrix with singular value decomposition $A=\sum_{i=0}^{n-1} \sigma_i|u_i\rangle\langle v_i|$. Then there is a quantum algorithm that runs in $O(\emph{poly}\log(n)/\epsilon)$ and achieves $\sum_{i=0}^{n-1} \alpha_i|v_i\rangle|0\rangle\mapsto\sum_{i=0}^{n-1}\alpha_i|u_i\rangle|\tilde{\sigma}_i\rangle$, where $|\tilde{\sigma}_i-\sigma_i|\leq \epsilon\|A\|_F$ for all $i$ with probability at least $1-1/\emph{poly}(n).$
\ep

\bo
Denote the norm of $w_i^\pm$ as $m_i^\pm$, the corresponding quantum states as $|w_i^\pm\rangle$. Since the eigenvalues and eigenvectors of $W$ contain the information of singular value and singular vectors of $A$, these information can be obtained by performing quantum phase estimation on $W$. The desired procedure can be obtained from the following five steps:

Step 1, choose the initial state as $\sum_{i=0}^{n-1} \alpha_i|v_i\rangle$, then apply $U_\mathcal{N}$ on it
\[
\sum_{i=0}^{n-1} \alpha_i\mathcal{N}|v_i\rangle=\sum_{i=0}^{n-1} \frac{\alpha_i}{2\i \sin(\theta_i/2)} \Big(m_i^+|w^+\rangle-m_i^-|w^-\rangle\Big).
\]

Step 2, perform the quantum phase estimation algorithm to estimate the eigenvalues and eigenvectors of $W$, then we get the following state
\[
\sum_{i=0}^{n-1} \frac{\alpha_i}{2\i \sin(\theta_i/2)}\Big(m_i^+|w^+\rangle|\theta_i\rangle-m_i^-|w^-\rangle|-\theta_i\rangle\Big).
\]

Step 3, change the phase and store the singular values in another register
\[
\ds\sum_{i=0}^{n-1} \frac{\alpha_i}{2\i \sin(\theta_i/2)}
\Big(e^{\i\theta_i/2}m_i^+|w^+\rangle|\theta_i\rangle-e^{-\i\theta_i/2}m_i^-|w^-\rangle|-\theta_i\rangle\Big)|\tilde{\sigma}_i\rangle.
\]

Step 4, undo the quantum phase estimation algorithm,
\[
\ds\sum_{i=0}^{n-1} \frac{\alpha_i}{2\i \sin(\theta_i/2)}\Big(e^{\i\theta_i/2}m_i^+|w^+\rangle-e^{-\i\theta_i/2}m_i^-|w^-\rangle\Big)|\tilde{\sigma}_i\rangle
=\sum_i\alpha_i\mathcal{M}|u_i\rangle|\sigma_i\rangle.
\]

Step 5, apply the inverse of $U_\mathcal{M}$ and we will get the desired state
$\sum_{i=0}^{n-1} \alpha_i|u_i\rangle|\tilde{\sigma}_i\rangle.$ The complexity mainly comes from the quantum phase estimation, which is $O(\textmd{poly}\log(n)/\epsilon)$.
\qed\eo

For any quantum state $|b\rangle=\sum_i\alpha_i|v_i\rangle$. To get the quantum information about $A|b\rangle$, in proposition \ref{svd}, when we obtain $\sum_i\alpha_i|u_i\rangle|\tilde{\sigma}_i\rangle$,
we can perform a controlled rotation on the register stores singular value and will get
\be\label{hhl-process}
\sum_{i=0}^{n-1}\alpha_i|u_i\rangle \Big(\tilde{\sigma}_i t|0\rangle+\sqrt{1-t^2\tilde{\sigma}_i^2}|1\rangle\Big),
\ee
where $t=1/\max_i\tilde{\sigma}_i$. By choosing a suitable $\epsilon$, we will get a good approximate of $|Ab\rangle$. As for our problem of computing the quantum state of $AB$, we can choose the initial state as $|B\rangle$, and implement the above procedure in parallel in each column. Finally by a simple analysis about the error and complexity (details are given in appendix \ref{app2}), we will get the following result

\bt \label{result2}
The quantum algorithm to get the quantum state of $AB$ to precision $\epsilon$ costs $\widetilde{O}(\|A\|_F\|B\|_F\kappa^2/\epsilon\|AB\|_F)$ by SVE.
\et

\br\label{remark-sve}
If $B=|b\rangle$, which only contains one column, then the complexity to obtain $|Ab\rangle$ is
$\widetilde{O}(\|A\|_F\kappa^2\epsilon\|A|b\rangle\|_2)
=\widetilde{O}(\sqrt{n}\kappa^3/\epsilon)$. However, to get a good approximate of $A|b\rangle$ without normalization, the procedure may not so expensive.
Actually, from (\ref{hhl-process}), we see that the error between $\sum_{i=0}^{n-1}\alpha_i\tilde{\sigma}_i|u_i\rangle$ and $A|b\rangle$ is bounded by $\epsilon\|A\|_F$. So we just need to choose $\epsilon\|A\|_F=\epsilon_1$ small. Then the complexity to get $A|b\rangle$ is $\widetilde{O}(\|A\|_F/\epsilon_1)$.
\er

The quantum algorithm to get the quantum state of $AB$ by HHL is similar to the algorithm by SVE.
We now assume that $A$ is Hermitian, otherwise we can consider
$\widetilde{A}=\left(
   \begin{array}{cc}
     0 & A \\
     A^\dagger & 0 \\
   \end{array}
 \right)$. We also assume that the Hamiltonian simulation of $e^{-\i \widetilde{A}t}$ is efficiently.
For any quantum state $|b\rangle=\sum\alpha_j|u_j\rangle$, by HHL algorithm, we can get the state $\alpha_j|u_j\rangle|\tilde{\sigma}\rangle$,
where $|\tilde{\sigma}_i-\sigma_i|\leq \epsilon \max_i\sigma_i$. Similarly, we can get (\ref{hhl-process}). Compared to the singular value estimation to achieve the quantum state of $|AB\rangle$, the only change is $\|A\|_F$, now it becomes $\max_i\sigma_i$. So by HHL algorithm, we can get the quantum state of $AB$ in time $\widetilde{O}(\kappa^3/\epsilon)$ to accuracy $\epsilon$, since $\|AB\|_F \geq \|B\|_F \min_i \sigma_i$.

\br \label{remark:SVE and HHL-other assumptions}
Just like HHL algorithm to solve linear system, the methods based on SVE and HHL also have the problem. More precisely, we potentially assume that each $|B_{\bullet j}\rangle$ lies in the nonzero components of $A$, i.e., the space generated by singular vectors with nonzero singular values. Otherwise, the success probability and so the final complexity will be affected.
For instance, in \eqref{hhl-process}, the success probability is $P=\sum_{i,\tilde{\sigma}_i\neq 0} |\alpha_i \tilde{\sigma}_i t|^2$. If we assume that $|b\rangle$ lies in the nonzero components, then $P\geq \min_{i,\tilde{\sigma}_i\neq 0} |\tilde{\sigma}_it|^2 \sum_{i,\tilde{\sigma}_i\neq 0} |\alpha_i|^2 =1/\kappa^2$, since $ \sum_{i,\tilde{\sigma}_i\neq 0} |\alpha_i|^2 =1$.
If there exists $i$ such that $\tilde{\sigma}_i\neq 0$ but $\alpha_i=0$, then $P\geq 1/\kappa^2$ may not hold anymore. So in these two algorithms, we should make this as another assumption.
However, the method based on swap test do not contain such a problem.
\er

The following table is a summary about the three quantum algorithms proposed in this section to achieve the quantum data of the multiplication of two matrices.

{\renewcommand\arraystretch{1.7}
\begin{table}[H]
\centering
\caption{Comparison of different quantum algorithms to achieve matrix multiplication with quantum information, where $A,B$ are input matrices, $\kappa$ is the condition number of $A$.}
\begin{tabular}{c|c|c}
  \hline \hline
   \hspace{.5cm}{\bf Methods} \hspace{.5cm} & \hspace{2.2cm} {\bf Complexity}  \hspace{2.2cm} & \hspace{2.9cm} {\bf Assumptions}  \hspace{2.9cm} \\\hline
  By swap test & $\widetilde{O}\left({\|A\|_F\|B\|_F}/{\epsilon\|AB\|_F}\right)=\widetilde{O}(\sqrt{n}\kappa/\epsilon)$ & \multirow{2}{*}{Efficient preparation of quantum states of $A$ and $B$} \\
  By SVE & $\widetilde{O}(\|A\|_F\|B\|_F\kappa^2/\epsilon\|AB\|_F)=\widetilde{O}(\sqrt{n}\kappa^3/\epsilon)$ &  \\\hline
  \multirow{2}{*}{By HHL} & \multirow{2}{*}{$\widetilde{O}(\kappa^3/\epsilon)$}
  & Efficient preparation of quantum states of $B$  \\
  && and efficient Hamiltonian simulation of $\widetilde{A}$ \\\hline\hline
\end{tabular} \label{quantum-data}
\end{table}}

Note that swap test and SVE work for all the cases if $A,B$ are classical or quantum data, while HHL needs $A$ or $B$ to be classical.
All the results are related to the condition number in the worst case. Simple analysis shows that $\sqrt{n}/\kappa \leq \|A\|_F\|B\|_F / \|AB\|_F \leq \sqrt{n} \kappa$, so if $\kappa=O(\textmd{poly}\log n)$, then $\|A\|_F\|B\|_F / \|AB\|_F=\widetilde{O}(\sqrt{n})$, however, the result by HHL is the best in this case. On the contrary, the complexities are not easy to determine approximately if $\kappa=\widetilde{O}(n^c)$ for some constant $c$.

\section{Quantum algorithm to matrix multiplication with classical information}
\setcounter{equation}{0}

In this section, we consider the problem of getting classical data to the multiplication of two matrices. First, we focus on the analysis of the method based on swap test. The quantum algorithms based on SVE or HHL are similar to analyze.

Let $\x,\y$ are two $n$ dimensional vectors. Denote the corresponding quantum states of these two vectors as $|x\rangle,|y\rangle$.
Then $\x\cdot\y=\|\x\|_2\|\y\|_2\langle x|y\rangle$, where $\|\x\|_2,\|\y\|_2$ are the norms of $\x,\y$.
By proposition \ref{cor:inner product}, we can get a good approximate of $\langle x|y\rangle$, i.e.,
we can get a value $P_{xy}$ in time $O(T_{\textmd{in}}/\epsilon)$ such that $|P_{xy}-\langle x|y\rangle|\leq \epsilon$.
However, a good approximate of $\langle x|y\rangle$ does not imply a good approximate of $\x\cdot\y$. This is because
$|P_{xy}\|\x\|_2\|\y\|_2-\x\cdot\y|\leq |\x||\y|\epsilon$. In order to make this error small, we denote
$\|\x\|_2\|\y\|_2\epsilon=\tilde{\epsilon}$, then
the final complexity of estimating $\x\cdot\y$ becomes
\be\label{complexity:inner product}
O(T_{\textmd{in}}\|\x\|_2\|\y\|_2/\tilde{\epsilon}).
\ee
Here we did not considered the complexity of evaluating $\|\x\|_2, \|\y\|_2$.
From the above analysis, we conclude that proposition \ref{cor:inner product} solves the inner product problem of two classical vectors efficiently only if the norms of
the vectors are small.

\br
We should remark that the influence of norms $\|\x\|_2, \|\y\|_2$ on the complexity (\ref{complexity:inner product})
by swap test to estimate inner product of $\x$ and $\y$ cannot removed actually.
This is all because of the optimality of Grover searching algorithm.
Consider the searching problem in $\mathbb{Z}_n$. Assume there are $r$ marked items, and $f:\mathbb{Z}_n\rightarrow\mathbb{Z}_2$ is defined as $f(i)=1$ if and only if $i$ is marked.
Now we define $g(x)=(-1)^{f(x)}$. Denote
$\x=(g(0),\ldots,g(n-1))$ and $\y=(1,\ldots,1).$
Then $\|\x\|_2=\|\y\|_2=\sqrt{n}$, and the quantum state $|x\rangle,|y\rangle$ can be prepared efficiently. As we can see $\x\cdot\y=\sum_{x=0}^{n-1} g(x)=n-2r$.
Suppose the complexity of evaluating the inner product of $\x\cdot\y$ is independent of $\|\x\|_2, \|\y\|_2$ and can be improved into $\widetilde{O}(1/\tilde{\epsilon})$,
then we can decide whether or not there exist marked items in $\mathbb{Z}_n$ efficiently, since we can just choose $\tilde{\epsilon}=O(1)$.
Together with the bisection method, we can finally find one marked item if $r>0$ efficiently. This will contradict the optimality of Grover searching algorithm.
\er

Let $A,B$ be two $n\times n$ matrices. Multiplying $A$ and $B$ is equivalent to evaluate $n^2$ inner product of $n$ dimensional vectors.
Classical method to evaluate inner product of two $n$ dimensional vectors takes time $O(n)$,
which lead the complexity of the classical matrix multiplication to $O(n^3)$.
However, swap test may reduce the complexity of evaluating inner product and so may reduce the complexity of matrix multiplication.
The norms of $A_{i\bullet},B_{\bullet j}$ can be evaluated by the classical method ($0\leq i,j\leq n-1$), which costs $O(n^2)$. These are classical data, and so can be used as many times as we want.
Since we assume that the quantum states of $A_{i\bullet},B_{\bullet j}$ can be prepared efficiently. Then by (\ref{complexity:inner product}) and note that
$A_{F\bullet},B_{\bullet F}$ are column vectors store the information of the 2 norms $A_{i\bullet},B_{\bullet j}$ (see remark \ref{remark-input-data}), we have

\bt \label{result1:classical}
There is a quantum algorithm that computes the multiplication of $A$ and $B$ with classical information in time
$
\widetilde{O}(\|A_{F\bullet}\|_1 \|B_{\bullet F}\|_1/\epsilon+n^2)
$
to accuracy $\epsilon$.
\et
\bo
From (\ref{complexity:inner product}), we know that the complexity to multiply $A,B$ with classical data is
$\sum_{i,j}\|A_{i\bullet}\|_2\|B_{\bullet j}\|_2/\epsilon=\|A_{F\bullet}\|_1 \|B_{\bullet F}\|_1/\epsilon$.
Together with $O(n^2)$ to compute the corresponding norms of $A_{i\bullet},B_{\bullet j}$, we will get the desired result.
\qed\eo

The accuracy $\epsilon$ in the theorem means that if $C=AB=(c_{ij})$ is the exact result and $\widetilde{C}=(\tilde{c}_{ij})$ is the result obtained from the quantum algorithm, then $|c_{ij}-\tilde{c}_{ij}|\leq \epsilon$. This is the absolute error. It is not easy to do the analysis of relative error now, since we have no more information about the value of inner product $\x\cdot \y$. However, by choosing the absolute error relatively smaller then the $\|\x\|_2\|\y\|_2$, the absolute error becomes closer to relative error.

Next, we consider the algorithms based on SVE and HHL.
In order to compute the classical information about $AB$, we actually do not need to perform measurements in the quantum algorithm by SVE or HHL. As discussed in remark \ref{remark-sve}, if $|B_j\rangle=\sum_k \alpha_{jk}|u_k\rangle$, then (\ref{hhl-process}) can be written as
\[
\frac{1}{\max_k\sigma_k}A|B_j\rangle|0\rangle+\textmd{orthogonal~part}.
\]
The above quantum state can be obtained efficiently in time $\widetilde{O}(1/\|A\|_F\epsilon)$.
By applying swap test on the above state with $|i,0\rangle$, we will get an approximate about the entries of $AB$. More details about the analysis of error and complexity, which are not so difficult, are given in appendix \ref{app3}. The final result is

\bt \label{result2:classical}
There is a quantum algorithm that computes the multiplication of $A$ and $B$ with classical information in time
$
\widetilde{O}({\sqrt{n} \kappa \|A_{F\bullet}\|^2\|B_{\bullet F}\|_3^3}/{\epsilon^2} +n^2)
$
by SVE and
$
\widetilde{O}({\kappa^2 \|A\|_F^2 \|B_{\bullet F}\|_3^3 }/{\epsilon_3^2} + n^2)
$
by HHL algorithm to accuracy $\epsilon$.
\et

The following table summarizes the above results about quantum matrix multiplication with classical data. Different from the algorithm to obtain quantum information, the complexity now is independent of $\|AB\|_F$. This is because that no measurements are needed in evaluating the classical data.

{\renewcommand\arraystretch{1.7}
\begin{table}[H]
\centering\caption{Comparison of different quantum algorithms to achieve matrix multiplication with classical information, where $A,B$ are input matrices, $\kappa$ is the condition number of $A$. }
\begin{tabular}{c|c|c}
  \hline\hline
   \hspace{.5cm}{\bf Methods} \hspace{.5cm} & \hspace{4cm} {\bf Complexity}  \hspace{4cm} & \hspace{1cm} {\bf Assumptions}  \hspace{1cm} \\\hline
  By swap test & $\widetilde{O}(\|A_{F\bullet}\|_1 \|B_{\bullet F}\|_1/\epsilon+n^2)
  =\widetilde{O}(n\|A\|_F \|B\|_F/\epsilon+n^2)$
  & \multirow{3}{*}{The same as table \ref{quantum-data}} \\
  By SVE       & $\widetilde{O}({\sqrt{n} \kappa \|A\|_F^2\|B_{\bullet F}\|_3^3}/{\epsilon^2} +n^2)
  =\widetilde{O}({\sqrt{n} \kappa \|A\|_F^2\|B\|_F^{3/2}}/{\epsilon^2} +n^2)
  $ & \\
  By HHL       & $\widetilde{O}({\kappa^2 \|A\|_F^2 \|B_{\bullet F}\|_3^3 }/{\epsilon^2} + n^2)
  =\widetilde{O}({\kappa^2 \|A\|_F^2 \|B\|_F^{3/2} }/{\epsilon^2} + n^2)$ & \\\hline\hline
\end{tabular}
\label{classical-data}
\end{table}}

To make the above table more easy to understand and easy to compare with classical algorithms, we assume that the singular values of $A,B$ are smaller than 1, then $\|A\|_F,\|B\|_F\leq \sqrt{n}$. So the above table can be simplified into

{\renewcommand\arraystretch{1.7}
\begin{table}[H]
\centering
\caption{Comparison of different quantum algorithms to achieve matrix multiplication with classical information, where $A,B$ are input matrices, $\kappa$ is the condition number of $A$ and the singular values of $A,B$ are smaller than 1. }
\begin{tabular}{c|c|c}
  \hline\hline
   \hspace{.6cm}{\bf Methods} \hspace{.6cm} & \hspace{.7cm} {\bf Complexity}  \hspace{.7cm} & \hspace{4.3cm} {\bf Assumptions}  \hspace{4.3cm} \\\hline
  By swap test & $\widetilde{O}(n^2/\epsilon)$
  & \multirow{3}{*}{Assumptions in table \ref{quantum-data} and the singular values of $A,B$ are smaller than 1} \\
  By SVE       & $\widetilde{O}(\kappa n^{2.25}/{\epsilon^2})$ & \\
  By HHL       & $\widetilde{O}({\kappa^2n^{1.75} }/{\epsilon^2} + n^2)$ & \\\hline\hline
\end{tabular}
\label{classical-data-simplified}
\end{table}}

The best classical algorithm to matrix multiplication with complexity $O(n^{2.3728639})$ is due to Le Gall \cite{legall} at 2014. If the precision
$\epsilon$ is small in size $O(1/\textmd{poly}\log n)$, then the quantum algorithm to matrix multiplication based on swap test is $\widetilde{O}(n^2)$. Also, to make this quantum algorithm better than Le Gall's classical algorithm, the upper bound of $\epsilon$ is
$O(n^{0.3728639}/\textmd{poly}\log n)$. The quantum algorithm based on SVE works better than the classical algorithm only if  $\kappa/\epsilon^2 = \widetilde{O}(n^{0.1228639})$ and the quantum algorithm based on HHL works better only if $\kappa^2/\epsilon^2= \widetilde{O}(n^{0.6228639})$. In \cite{buhrman-matrix}, Buhrman et al also proposed a quantum algorithm to achieve matrix multiplication, however, their complexity depends on the number of nonzero entries of $AB$, so we prefer not to compare with it here.

\section{Preparation of quantum states}
\setcounter{equation}{0}
\label{Preparation of quantum states}

Let $\x=(x_0,\ldots,x_{n-1})$ be a complex vector, the quantum state it corresponds to equals $|x\rangle=\frac{1}{\|\x\|_2} \sum_{i=0}^{n-1} x_i|i\rangle$. The transformation from classical data $\x$ into its quantum state is usually called ``input problem" in quantum computer, which plays important roles in many quantum algorithms, such as \cite{childs-linear-systems}, \cite{clader}, \cite{harrow}, \cite{kerenidis}, \cite{kerenidis-iteration}, \cite{lloyd}, \cite{rebentros}, \cite{rebentros-newton}, \cite{rebentros-svd}, \cite{wang}, \cite{wiebe}, \cite{wossnig}. Moreover, the quantum matrix multiplication algorithms proposed above also rely on efficient preparation of quantum states. This section is devoted to study the input problem.

The most naive method is defining a unitary $U$ such that $U|0\rangle=|x\rangle$. The efficiency of preparing $|x\rangle$ is totally determined by $U$. In the worst case, $U$ can be implemented in time $O(n^2(\log n)^2\log^c(n^2(\log n)^2/\epsilon))$ (see \cite[Chapter 4]{nielsen}) to precision $\epsilon$ in quantum computer, where $c$ is some constant close to 2. So we can prepare $|x\rangle$ within the same time in the worst case. Conclude this, we have

\bp \label{state prepare: direct method}
For any vector $\x$, its quantum state can be prepared in time $O(n^2(\log n)^2\log^c(n^2(\log n)^2/\epsilon))$  to precision $\epsilon$.
\ep

Although the above method works for all cases, it is not efficient generally. We still hope the input problem can be solved efficiently in some special cases. Under certain conditions, this problem can actually solved efficiently \cite{clader}, \cite{grover-state-preparation}, \cite{lloyd}, \cite{soklakov}. In the following, we focus on the one given in \cite{lloyd}.
In this paper, Lloyd et al provided a quantum algorithm to prepare quantum state, which works very well when the given classical data are relatively uniform distributed. It has been used to solve the supervised classification problem \cite{lloyd} and the least square support vector machine problem \cite{rebentros}.
In the following, we first give a detailed analysis about this technique in order to find its advantages and disadvantages. Then based on this algorithm, we will propose two new quantum algorithms with better efficiency.

Let $f$ be a map from $\mathbb{Z}_n$ to $\mathbb{R}$, denote
\[
\ds\max(f):=\max_{k\in\mathbb{Z}_n} |f(k)|;~~~~
\min(f):=\min_{\ k\in\mathbb{Z}_n, f(k)\neq 0} |f(k)|;~~~~
\kappa(f):={\max(f)}/{\min(f)}.
\]
These notations can be similarly extended to vectors or sequences. We should remark that in \cite{lloyd}, the authors's main objective is solving the supervised classification problem, so their results and methods are confined to this problem. Moreover, they did not give too much analysis about the efficiency of their method. However, their method is more general then preparing quantum states. The following result is obtained by generalizing their method.

\bp \label{prop:state prepartion}
Let $f$ be a map (or an oracle) from $\mathbb{Z}_n$ to $\mathbb{R}^*=\mathbb{R}\setminus\{0\}$.
Then for any state $|\psi\rangle=\sum_{k=0}^{n-1}b_k|k\rangle$ with preparation complexity $O(T_{|\psi\rangle})$, we can construct the state $|\psi'\rangle=\frac{1}{\sqrt{Z}}\sum_{k=0}^{n-1}f(k)b_k|k\rangle$ in time
$
O({ \kappa(f)^{3/2}}T_{|\psi\rangle}/{\epsilon})
$
to accuracy $\epsilon$, where $Z=\sum_{k=0}^{n-1} |f(k)b_k|^2$.
\ep

\bo
Let $H=\sum_{k=0}^{n-1}f(k)|k\rangle\langle k|$ be a Hamiltonian, which is a diagonal matrix, so $e^{-\i Ht}$ can be implemented efficiently.
Consider the following procedure:
\be\ba{lll} \vspace{.2cm} \label{alg: state prepartion}
\ds\frac{1}{\sqrt{2}}(|0\rangle+|1\rangle)\sum_{k=0}^{n-1}b_k|k\rangle
&\mapsto&
\ds\frac{1}{\sqrt{2}}|0\rangle\sum_{k=0}^{n-1}b_ke^{\i f(k)t}|k\rangle+\frac{1}{\sqrt{2}}|1\rangle\sum_{k=0}^{n-1}b_ke^{-\i f(k)t}|k\rangle \\ \vspace{.2cm}
&  \mapsto&
\ds\frac{1}{\sqrt{2}}|+\rangle\sum_{k=0}^{n-1}b_ke^{\i f(k)t}|k\rangle+\frac{1}{\sqrt{2}}|-\rangle\sum_{k=0}^{n-1}b_ke^{-\i f(k)t}|k\rangle \\
&  =& \ds|0\rangle\sum_{k=0}^{n-1}b_k\cos(f(k)t)|k\rangle+\i|1\rangle\sum_{k=0}^{n-1}b_k\sin(f(k)t)|k\rangle,
\ea\ee
where the first step is the result of Hamiltonian simulation and the second step applies Hadamard transformation on the first qubit.

Choosing $t$ small enough such that there exist $\epsilon_0,\epsilon_1$ satisfy $\epsilon_0\leq |f(k)t|\leq \epsilon_1\ll1$,
here we choose $\epsilon_1=O(\kappa(f)^{-1/2}\epsilon)$. Then the state along with $|1\rangle$ is an approximation of the state $|\psi'\rangle$.
And the error between them is bounded by $O(\epsilon)$ (see more details in appendix \ref{app4}).
Note that $f(k)\neq0$ for all $k$, so the probability of getting $|1\rangle$ is
$P\approx Zt^2=\sum_{k=0}^{n-1}|b_kf(k)t|^2\geq\epsilon_0^2.$
By amplitude amplification technique, it suffices measuring $O(1/\epsilon_0)$ times.
Note that $\epsilon_0\leq |f(k)t|\leq \epsilon_1$, so we have $\epsilon_0\approx\min(f)t$ and $\epsilon_1\approx\max(f)t$.
Hence
$\epsilon_0=\kappa(f)^{-1}\epsilon_1=O(\kappa(f)^{-3/2}\epsilon).$
The complexity of procedure (\ref{alg: state prepartion}) is $O(T_{|\psi\rangle})$, so the complexity of getting $|\psi'\rangle$ is
$O({T_{|\psi\rangle}}/{\epsilon_0})=O({\kappa(f)^{3/2}}T_{|\psi\rangle}/{\epsilon})$.
\qed\eo

The final complexity of the algorithm given in proposition \ref{prop:state prepartion} is affected by $\kappa(f)$. A simple case is when the sequence $\{|f(0)|,\cdots,|f(n-1)|\}$ is relatively uniform. Here relatively uniform means $\kappa(f)=\max(f)/\min(f)$ is an acceptable small constant. In this case, the complexity of proposition \ref{prop:state prepartion} can be further simplified into
$O(T_{|\psi\rangle}/\epsilon).$

In \cite{clader}, Clader et al also propose a quantum algorithm to prepare quantum state, which is indirectly inspired by the work of  HHL algorithm \cite{harrow}. Simple analysis shows that their algorithm also contains the same problem discussed above, that is the influence of $\kappa(f)$. However, their algorithm is quite simple and contains no error.
Note that the idea of proposition \ref{prop:state prepartion} can be generalized into the complex field by considering the real and image part respectively, so in the following we just need to focus on the preparation of real vectors.
Moreover, if there exists $k$ such that $f(k)=0$, then the probability analysis
$P=\sum_{k=0}^{n-1}|b_kf(k)t|^2\geq\epsilon_0^2$ may not hold anymore.
However, if $|\psi\rangle=\frac{1}{\sqrt{n}}\sum_{k=0}^{n-1}|k\rangle$ and the sequence $\{|f(0)|,\cdots,|f(n-1)|\}$
contains $O(z)$ nonzero elements, then the complexity of proposition \ref{prop:state prepartion} should be multiplied by $O(\sqrt{n/z})$ because of amplitude amplification technique.
A direct application of proposition \ref{prop:state prepartion} is

\bp
Let $\x=(x_0,\ldots,x_{n-1})$ be a given vector that contains $O(z)$ nonzero elements, denote $\kappa(\x)={\max_k |x_k|}/{\min_{k,x_k\neq 0} |x_k|}$,
then  $|x\rangle=\frac{1}{\|\x\|_2}\sum_{k=0}^{n-1}x_k|k\rangle$
can be prepared in time
$O({\kappa(\x)^{3/2}\sqrt{n/z}(\log n) }/{\epsilon})$
to accuracy $\epsilon$. Moreover, if $\x$ is relatively uniform, then the complexity is
$O(\sqrt{n/z}(\log n)/\epsilon).$
\ep

\bo Just choose $|\psi\rangle=\frac{1}{\sqrt{n}}\sum_{k=0}^{n-1}|k\rangle$ and $f(k)=x_k$ in proposition \ref{prop:state prepartion}.
\qed\eo

However, if we know the positions of nonzero entries in $\x$, then we can just focus on these nonzero parts and apply the algorithm given in proposition \ref{prop:state prepartion} to prepare the quantum state of the nonzero components of $\x$, which is also equals to the quantum state of $\x$ itself. So we have

\bp \label{state prepare: indirect method 1}
For any given vector $\x$, there is a quantum algorithm to prepare its quantum state in time
$O(\kappa(\x)^{3/2}{(\log n) }{\epsilon})$. Moreover, if $\x$ is relatively uniform, then the complexity is $O((\log n)/\epsilon)$ .
\ep

As we can see from the above result, the quantum state preparation algorithm works efficient when $\x$ is a relatively uniform distributed vector, and may performs very bad otherwise. One way to grasp the property of relatively uniform distribution is decomposing $\x$ into the sum of several relatively uniform distributed vectors. In the following, we given two different such decompositions.

Let $\x=(x_0,\ldots,x_{n-1})$ be a real vector. Assume that $0<x_0<\cdots<x_{n-1}$ for simplicity. Denote $\kappa(\x)=x_{n-1}/x_0\approx 2^q$. Then each interval $I_j=[2^{j-1}x_0,2^{j}x_0)$, where $1\leq j\leq q$, contains several values of $x_i$, we denote them as $x_{j1},\ldots,x_{jt_j}$ and set the vector $\y_j=(0,\ldots,0,x_{j1},\ldots,x_{jt_j},0,\ldots,0)$. Then $\x=\y_1+\cdots+\y_q$ and each vector $\y_j$ can be prepared efficiently in time $O((\log n)/\epsilon)$ by proposition \ref{state prepare: indirect method 1}. Now we see that $|x\rangle=\lambda_1|y_1\rangle+\cdots+\lambda_q|y_q\rangle$, where $\lambda_j=\|\y_j\|_2/\|\x\|_2$. From the method given in \cite[Chapter 26]{childs}, the complexity to achieve such a linear combination to get $|x\rangle$ equals
$
O(C_q(\log n)\sum_{j=1}^q \|\y_j\|_2/\|\x\|_2\epsilon)=O(q^{5/2}(\log q)^2\log^c(q^2(\log q)^2/\epsilon)(\log n)/\epsilon),
$
where $C_q$ is the complexity to implement the unitary $U$ such that $U|0\rangle \propto\sum \sqrt{\lambda_j}|j\rangle$, which is at most
$O(q^2(\log q)^2\log^c(q^2(\log q)^2/\epsilon))$  as discussed in \cite[Chapter 4]{nielsen}. When the entries of $\x$ are not all positive and increasing, we define the interval $I_j$ based on the vector whose entries are the absolute value of the entries of $\x$ and nonzero. The requirement that all entries are sorted is not so necessary, since the above analysis are also hold for the case that $\x$ is not sorted, we only need to focus on positions of vectors lie in $I_j$. At this case, the notations will be a little complicate, but it changes nothing. Therefore, we have

\bt \label{state prepare: indirect method 2}
Let $\x=(x_0,\ldots,x_{n-1})$ be a given vector and $\kappa(\x)=\max_k |x_k|/\min_{k,x_k\neq 0} |x_k|$. Then its quantum state can be prepared in time $O((\log \kappa(\x))^{5/2}(\log \log \kappa(\x))^2\log^c[(\log \kappa(\x))^2(\log \log \kappa(\x))^2/\epsilon](\log n)/\epsilon)$.
\et

This result is better than proposition \ref{state prepare: indirect method 1}. If $\kappa(\x)$ is too large, then we may consider giving up the components that are close to $\min_{k,x_k\neq 0} |x_k|$ if they are not too many. Moreover, even if $\max|x_k|=2^{1000}$ and $\min_{k,x_k\neq 0} |x_k|=1$, then $\log \kappa(\x)=1000$, which is still an acceptable constant. From this point, the above result which is polynomially depending on $\log \kappa(\x)$ seems to be a pretty good algorithm to prepare quantum states.

Another decomposition is more direct and easy. We assume that all entries of $\x$ are nonzero, otherwise, we only focus on the nonzero components.
Now define $\y=M(\textmd{sign}(x_0),\ldots,\textmd{sign}(x_{n-1}))$, where $M\geq\max_i|x_i|$ and $\textmd{sign}(x_i)=1$ if $x_i\geq 0$; $\textmd{sign}(x_i)=-1$ if $x_i < 0$.
Then the quantum state $|y\rangle$ of $\y$ can be obtained efficiently in time $O(\log n)$. Also define $\z:=\x+\y
=(\textmd{sign}(x_0)M+x_0,\ldots,\textmd{sign}(x_{n-1})M+x_{n-1})$, which is uniformly distributed with $n$ nonzero entries. So by proposition \ref{state prepare: indirect method 1}, the quantum state $|z\rangle$ of $\z$ can be obtained efficiently in time $O((\log n)/\epsilon)$ to precision $\epsilon$. Since
\[
|x\rangle = \frac{1}{\|\x\|_2} (\z-\y) = \frac{\|\z\|_2}{\|\x\|_2}|z\rangle - \frac{\|\y\|_2}{\|\x\|_2}|y\rangle.
\]
What we should do next is computing the linear combination of two efficiently prepared quantum states. The linear combination of two quantum states can be obtained from a similar procedure to Hadamard test as follows: here for simplicity denote $\|\z\|_2/\|\x\|_2$ as $\lambda$, denote $\|\y\|_2/\|\x\|_2$ as $\mu$,
\[\ba{lll} \vspace{.2cm}
\ds\frac{1}{\sqrt{\lambda^2+\mu^2}} (\lambda |0\rangle |z\rangle+\mu |1\rangle |y\rangle)
&\mapsto&
\ds\frac{1}{\sqrt{\lambda^2+\mu^2}} (\lambda |+\rangle |z\rangle+\mu |-\rangle |y\rangle) \\
&=& \ds \frac{1}{\sqrt{2(\lambda^2+\mu^2)}} \Big[ |0\rangle (\lambda |z\rangle+\mu  |y\rangle)
+ |1\rangle (\lambda |z\rangle-\mu  |y\rangle)\Big].
\ea\]
The probability to get $|x\rangle=\lambda |z\rangle-\mu  |y\rangle$ is $1/2(\lambda^2+\mu^2)=\|\x\|_2^2/2(\|\y\|_2^2+\|\z\|_2^2)$. Then the complexity  to get $|x\rangle$ is
$O(\sqrt{(\|\y\|_2^2+\|\z\|_2^2)/\|\x\|_2^2}(\log n)/\epsilon)$. Set
\[
d=\frac{\max_k (|\textmd{sign}(x_k)M+x_k|)}{\min_k(|\textmd{sign}(x_k)M+x_k|)}=\frac{M+\max_k|x_k|}{M+\min_k|x_k|}.
\]
If $d$ is a small constant, then we can just choose $M=\max_k|x_k|$. Note that $\|\x\|_2\geq n\min_k|x_k|$, so
\[
\frac{\|\y\|_2^2+\|\z\|_2^2}{\|\x\|_2^2}
=\frac{2nM^2+2M\sum_i x_i+\|\x\|_2^2}{\|\x\|_2^2}
\leq 3\kappa(\x)^2+1.
\]
Hence, the complexity obtained by this decomposition is $O(\kappa(\x)(\log n)/\epsilon)$, which is better by applying proposition \ref{state prepare: indirect method 1} directly, however, not efficient than theorem \ref{state prepare: indirect method 2}.
In \cite{shao}, the author propose a method to achieve the linear combination, which is independent of the effect of $\lambda$ and $\mu$,
in time $O(\log(n)/\epsilon^2)$. Therefore, we have

\bt \label{state prepare: indirect method 3}
For any vector $\x$, its quantum state can be prepared in time $O(\log(n)/\epsilon^2)$ to precision $\epsilon$.
\et

In the following, we summarize all the quantum algorithms proposed above in the following table. Note that they work for all classical data, so no other assumptions are needed. However, they need a lot of information about the input data, such as the nonzero components, the maximal and minimal entries. It may take some extra time to get them (such as by quantum searching algorithm \cite{durr-find-minimun}, \cite{grover}) and we will not consider it right here.

{\renewcommand\arraystretch{1.7}
\begin{table}[H]
\centering
\caption{Comparison of different quantum algorithms to prepare quantum state of $\x=(x_0,\ldots,x_{n-1})$,
where  $\kappa(\x)=\max_k |x_k|/\min_{k,x_k\neq 0} |x_k|$ and $\epsilon$ is the precision.}
\begin{tabular}{c|c}
  \hline\hline
   \hspace{.9cm}{\bf Algorithms given in} \hspace{1cm} & \hspace{5cm} {\bf Complexity}  \hspace{5cm}  \\ \hline
   Proposition \ref{state prepare: direct method}       & $O(n\log(n)\log(1/\epsilon))$ \\
   Proposition \ref{state prepare: indirect method 1} & $O(\kappa(\x)^{3/2}{(\log n) }/{\epsilon})$ \\
   Theorem \ref{state prepare: indirect method 2} & $O((\log \kappa(\x))^{5/2}(\log \log \kappa(\x))^2\log^c[(\log \kappa(\x))^2(\log \log \kappa(\x))^2/\epsilon](\log n)/\epsilon) $ \\
   Theorem \ref{state prepare: indirect method 3} &  $O(\log(n)/\epsilon^2)$ \\ \hline \hline
\end{tabular}
\label{Comparison: quantum state preparation algorithms}
\end{table}}

In the quantum algorithms to achieve matrix multiplication with classical data, we can first apply searching algorithm to find the desired information to prepare quantum states, which takes at most $O(n^2)$ steps and does not affect the final complexity of the algorithms. This means, to getting the classical data of matrix multiplication, quantum algorithms listed in table \ref{classical-data}, except HHL, do not need the assumption listed in the beginning of section 2. Strictly speaking, actually only the quantum algorithm obtained by swap test do not need any assumptions by the analysis in remark \ref{remark:SVE and HHL-other assumptions}. Therefore, we have

\bt
The multiplication of two $n\times n$ matrices can be obtained in time $\widetilde{O}(n^2/\epsilon)$ to precision $\epsilon$.
\et

\section{Conclusions}

Quantum computer outperforms the classical computer in many problems. However, many of the quantum algorithms make one or two assumptions; the most common one is the input problem, that is we assume the given data is quantum data by some methods like QRAM. However, this problem is not easy to solve generally. As suggested in \cite{biamonte}, we can study quantum algorithms only with quantum input and output. With this idea, we do not need to consider the input and output problem.
One important task worth to study is extending classical matrix operations into quantum computer with quantum input and output data. In this paper, we only considered the problem of matrix multiplication, however, it forms the most elementary step of many other matrix operations, such as QR decomposition and LU decomposition. Until now, most matrix operations are not easy to find suitable quantum techniques to deal with them efficiently. QR decomposition is very useful, however, it seems quit difficult to make it efficient in quantum computer in polynomial time. Fortunately, we already have SVE technique and its generalized version. This will play important roles in studying quantum matrix operations. The generalized version of swap test can be viewed as another technique that we can apply.

\vspace{.3cm}

~\\
{\bf Acknowledgements.} This work is supported by the NSFC Project 11671388 and the CAS Frontier Key Project QYZDJ-SSW-SYS022.

\appendix

\section{Error and complexity analysis of theorem \ref{result1}}
\setcounter{equation}{0}
\label{app1}

For simplicity, we denote the approximate of $s_{ij}:=\langle A_{i\bullet}|B_{\bullet j}\rangle$ obtained from the algorithm as $\tilde{s}_{ij}$.
Then $|s_{ij}-\tilde{s}_{ij}|\leq \epsilon$. Denote
\[
|C\rangle=\frac{1}{\|C\|_F}\sum_{i,j=0}^{n-1} C_{ij}|i,j\rangle=\frac{1}{\|C\|_F}\sum_{i,j=0}^{n-1}\|A_{i\bullet}\|_2\|B_{\bullet j}\|_2s_{ij} |i,j\rangle,
\]
and
\[
|\widetilde{C}\rangle=\frac{1}{\sqrt{Z}}\sum_{i,j=0}^{n-1}\|A_{i\bullet}\|_2\|B_{\bullet j}\|_2\tilde{s}_{ij} |i,j\rangle
\]
as the quantum state obtain from the algorithm, where
\[
Z=\sum_{i,j=0}^{n-1}\|A_{i\bullet}\|_2^2\|B_{\bullet j}\|_2^2\tilde{s}_{ij}^2.
\]
Then
\[\ba{rll} \vspace{.2cm}
|\|C\|_F^2-Z| &=& \ds\sum_{i,j=0}^{n-1}\|A_{i\bullet}\|_2^2\|B_{\bullet j}\|_2^2|s_{ij}^2-\tilde{s}_{ij}^2|
\leq 2\epsilon\sum_{i,j=0}^{n-1}\|A_{i\bullet}\|_2^2\|B_{\bullet j}\|_2^2=2\epsilon\|A\|_F^2\|B\|_F^2; \\
|\|C\|_F-\sqrt{Z}| &=& \ds \frac{\|C\|_F^2-Z|}{\|C\|_F+\sqrt{Z}}\leq \frac{2\epsilon\|A\|_F^2\|B\|_F^2}{\|C\|_F+\sqrt{Z}}.
\ea\]
Finally,
\[\ba{lll} \vspace{.2cm}
\||C\rangle-|\widetilde{C}\rangle\|^2_2
&\leq& \ds\frac{1}{\|C\|_F^2Z}\sum_{i,j=0}^{n-1}\|A_{i\bullet}\|_2^2 \|B_{\bullet j}\|_2^2 \Big(s_{ij}\sqrt{Z}-\tilde{s}_{ij}\|C\|_F\Big)^2 \\ \vspace{.2cm}
&\leq& \ds\frac{2}{\|C\|_F^2Z}\sum_{i,j=0}^{n-1}\|A_{i\bullet}\|_2^2 \|B_{\bullet j}\|_2^2
       \Big(\tilde{s}_{ij}^2(\sqrt{Z}-\|C\|_F)^2+Z(s_{ij}-\tilde{s}_{ij})^2\Big) \\ \vspace{.2cm}
&\leq&\ds \frac{2\|A\|_F^2\|B\|_F^2}{\|C\|_F^2}\epsilon^2+\frac{8\|A\|_F^4\|B\|_F^4}{\|C\|_F^2(\|C\|_F+\sqrt{Z})^2}\epsilon^2 \\
&\leq&\ds \frac{2\|A\|_F^2\|B\|_F^2}{\|C\|_F^2}\epsilon^2+\frac{2\|A\|_F^4\|B\|_F^4}{\|C\|_F^4}\epsilon^2.
\ea\]

To make the error is small in size $\epsilon_0$, we need to choose
${\epsilon\|A\|_F^2\|B\|_F^2}/{\|C\|_F^2}=\epsilon_0$. Since the success probability is $Z/\|A\|_F^2\|B\|_F^2\approx\|C\|_F^2/\|A\|_F^2\|B\|_F^2$, finally the complexity will be $\widetilde{O}(\|A\|_F\|B\|_F/\|C\|_F\epsilon)=\widetilde{O}(\|A\|_F^3\|B\|_F^3/\|C\|_F^3\epsilon_0)$.

\section{Error and complexity analysis of theorem \ref{result2}}
\setcounter{equation}{0}
\label{app2}

Denote $|B_{\bullet j}\rangle=\sum_{k=0}^{n-1} \alpha_{jk}|u_k\rangle$, then similar to the procedure (\ref{hhl-process}), by choosing the initial state as
\[
|B\rangle=\frac{1}{\|B\|_F} \sum_{j=0}^{n-1} \|B_{\bullet j}\|_2 |B_{\bullet j}\rangle|j\rangle
=\frac{1}{\|B\|_F} \sum_{j,k=0}^{n-1} \|B_{\bullet j}\|_2 \alpha_{jk}|u_k\rangle|j\rangle,
\]
we can get the following state
\be\label{AB-state-by-sve}
\frac{1}{\|B\|_F\max_k\tilde{\sigma}_k} \sum_{j,k=0}^{n-1} \|B_{\bullet j}\|_2 \alpha_{jk}\tilde{\sigma}_k |u_k\rangle|j\rangle|0\rangle+\textmd{orthogonal~parts}.
\ee

Now denote
\[
|\phi\rangle = \frac{1}{\sqrt{Z}} \sum_{j,k=0}^{n-1} \|B_{\bullet j}\|_2 \alpha_{jk}\tilde{\sigma}_k |u_k\rangle|j\rangle, \hspace{1cm}
|\psi\rangle = \frac{1}{\sqrt{W}} \sum_{j,k=0}^{n-1} \|B_{\bullet j}\|_2 \alpha_{jk}\sigma_k |u_k\rangle|j\rangle,
\]
where $Z=\sum_{j,k=0}^{n-1} \|B_{\bullet j}\|_2^2 |\alpha_{jk}|^2 \tilde{\sigma}_k^2$ and
$W=\sum_{j,k=0}^{n-1} \|B_{\bullet j}\|_2^2 |\alpha_{jk}|^2 \sigma_k^2$. Since $|\tilde{\sigma}_k-\sigma_k|\leq \epsilon\|A\|_F$ and
$\|B\|_F^2=\sum_{j,k=0}^{n-1} \|B_{\bullet j}\|_2^2|\alpha_{jk}|^2$, we have
\[
|Z-W| \leq \sum_{j,k=0}^{n-1} \|B_{\bullet j}\|_2^2 |\alpha_{jk}|^2 |\tilde{\sigma}_k^2-\sigma_k^2|
\leq \epsilon\|A\|_F\|B\|_F^2 \max_{0\leq k\leq n-1} |\tilde{\sigma}_k+\sigma_k|.
\]

Therefore,
\be\ba{lll} \label{bound}\vspace{.2cm}
\|\phi\rangle-|\psi\rangle\|_2^2
&=& \ds\frac{1}{ZW} \Bigg\| \sum_{j,k=0}^{n-1} \|B_{\bullet j}\|_2 \alpha_{jk} (\sqrt{W}\tilde{\sigma}_k - \sqrt{Z}\sigma_k)|u_k\rangle|j\rangle   \Bigg\|_2^2 \\ \vspace{.2cm}
&\leq& \ds\frac{2}{ZW} \sum_{j,k=0}^{n-1} \|B_{\bullet j}\|_2^2 |\alpha_{jk}|^2 \Big(W(\tilde{\sigma}_k-\sigma_k)^2 + (\sqrt{Z}-\sqrt{W})^2\sigma_k^2\Big) \\
&\leq& \ds\frac{2\epsilon^2\|A\|_F^2\|B\|_F^2}{Z}+\frac{2\epsilon^2\|A\|_F^2\|B\|_F^4 \max_k |\tilde{\sigma}_k+\sigma_k|^2}{Z(\sqrt{Z}+\sqrt{W})^2}.
\ea\ee
By choosing $\epsilon_1=\epsilon\|A\|_F$ small, then $\tilde{\sigma}_k\approx\sigma_k$, so $|Z-W|\leq 2\epsilon_1\|B\|_F^2\max_k \sigma_k $.
Now choosing $\epsilon_2=\epsilon_1\|B\|_F^2\max_k \sigma_k $, we have $Z\approx W\leq \|B\|_F^2\max_k \sigma_k^2$, so the upper bound given in (\ref{bound}) is close to
\[\ba{lll} \vspace{.2cm}
\ds\frac{2\epsilon^2\|A\|_F^2\|B\|_F^2}{W}+\frac{\epsilon^2\|A\|_F^2\|B\|_F^4 \max_k \sigma_k^2}{W^2}
&=& \ds\frac{2\epsilon_1^2\|B\|_F^2}{\|AB\|_F^2}+\frac{\epsilon_2^2}{\|AB\|_F^4} \\ \vspace{.2cm}
&=& \ds\frac{\epsilon_2^2}{\|AB\|_F^2}\Big(  \frac{2}{\|B\|_F^2\max_k \sigma_k^2}+\frac{1}{\|AB\|_F^2}\Big) \\
&\leq&  \ds\frac{3\epsilon_2^2}{\|AB\|_F^4}.
\ea\]
Finally, we set $\epsilon_2=\epsilon_3\|AB\|_F^2$ which makes the error between $|\phi\rangle$ and $|\psi\rangle$ smaller than $\epsilon_3$. The probability of getting $|\phi\rangle$ in \eqref{AB-state-by-sve} equals $Z/\|B\|_F^2\max_k \tilde{\sigma}_k^2\approx \|AB\|_F^2/\|B\|_F^2\max_k \sigma_k^2$. So the complexity to get $|\phi\rangle$ is
\[
\frac{\|B\|_F\max_k \sigma_k}{\|AB\|_F\epsilon}
=\frac{\|A\|_F\|B\|_F^3\max_k \sigma_k^2}{\|AB\|_F^3\epsilon_3}
\leq \frac{\|A\|_F\|B\|_F\kappa^2}{\|AB\|_F\epsilon_3},
\]
since $\|AB\|_F^2\geq \|B\|_F^2 \min_k \sigma_k^2$.

\section{Error and complexity analysis of theorem \ref{result2:classical}}
\setcounter{equation}{0}
\label{app3}

Set $|B_{\bullet j}\rangle=\sum_{k=0}^{n-1} \alpha_{jk}|u_k\rangle$ and $\epsilon \|A\|_F=\epsilon_1$, then in \eqref{hhl-process}, we obtain
\[
\frac{1}{\max_k\sigma_k} \sum_k \alpha_{jk} \tilde{\sigma}_k|u_k\rangle|0\rangle
+\textmd{orthogonal~part}
\]
in time $\widetilde{O}(\|A\|_F/\epsilon_1)$. Apply swap test on the state $|i,0\rangle$ and the above state, we will get a value $L$ in time $\widetilde{O}(\|A\|_F/\epsilon_1\epsilon_2)$, such that
\[
\Bigg|L-\frac{1}{\|B_{\bullet j}\|_2\max_k\sigma_k} \sum_k \|B_{\bullet j}\|_2 \alpha_{jk} \tilde{\sigma}_k \langle i|u_k\rangle\Bigg|\leq \epsilon_2.
\]
Note that $c_{ij}= \sum_k\|B_{\bullet j}\| \alpha_{jk} \sigma_k \langle i |u_k\rangle$, so
\[\ba{lll} \vspace{.2cm}
\Big|L\|B_{\bullet j}\| \max_k \sigma_k-c_{ij}\Big|
&\leq&\ds\Big|L\|B_{\bullet j}\|_2 \max_k \sigma_k-\sum_k \|B_{\bullet j}\|_2 \alpha_{jk} \tilde{\sigma}_k\langle i|u_k\rangle\Big| \\\vspace{.2cm}
&& +\ds\Big|\sum_k\|B_{\bullet j}\|_2 \alpha_{jk} \sigma_k \langle i |u_k\rangle-\sum_k \|B_{\bullet j}\|_2 \alpha_{jk} \tilde{\sigma}_k\langle i|u_k\rangle\Big|  \\ \vspace{.2cm}
&\leq& \ds \epsilon_2\|B_{\bullet j}\|_2 \max_k \sigma_k
+\epsilon_1\|B_{\bullet j}\|_2 \sum_k|\alpha_{jk} \langle i|u_k\rangle|.
\ea\]

We choose $\epsilon_1$ and $\epsilon_2$ such that
$\epsilon_2\|B_{\bullet j}\|_2 \max_k \sigma_k=\epsilon_3$
and $\epsilon_1\|B_{\bullet j}\|_2 \sum_k|\alpha_{jk} \langle i|u_k\rangle|=\epsilon_3$.
Finally, the complexity is
\[
O\Bigg(\frac{\|A\|_F}{\epsilon_1\epsilon_2}\Bigg)
=O\Bigg(\frac{\|A\|_F\|B_{\bullet j}\|_2^2\max_k \sigma_k\sum_k|\alpha_{jk} \langle i|u_k\rangle| }{\epsilon_3^2}\Bigg).
\]
This is the complexity to compute $c_{ij}$ to accuracy $\epsilon_3$. Therefore, the total complexity to compute all entries of $C$ equals
\be\label{app3:eq}
O\Bigg( \frac{\|A\|_F\max_k \sigma_k }{\epsilon_3^2} \sum_{i,j,k}\|B_{\bullet j}\|^2|\alpha_{jk} \langle i|u_k\rangle| +n^2\Bigg)
=O\Bigg(\frac{n\|A\|_F\|B_{\bullet F}\|_3^3\max_k \sigma_k}{\epsilon_3^2} +n^2\Bigg),
\ee
which is due to
\[
\sum_{i,j,k}\|B_{\bullet j}\|_2^2|\alpha_{jk}| |\langle i|u_k\rangle|
=\sum_{j,k}\|B_{\bullet j}\|_2^2|\alpha_{jk}| \|u_k\|_1 \leq \sqrt{n}\sum_{j}\|B_{\bullet j}\|_2^2 \sum_k|\alpha_{jk}|
= \sqrt{n}\sum_{j}\|B_{\bullet j}\|_2^2 \|B_{\bullet j}\|_1
\leq n\sum_{j}\|B_{\bullet j}\|^3.
\]
Since $\|A\|_F\geq \sqrt{n} \min_k \sigma_k$, the above result can be changed into
\[
O\Bigg(\frac{\sqrt{n} \kappa \|A\|_F^2\|B_{\bullet F}\|_3^3}{\epsilon_3^2} +n^2\Bigg).
\]

If we apply HHL algorithm, then $\|A_{F\bullet}\|$ should be changed into $\max_k \sigma_k$, so the complexity in (\ref{app3:eq}) becomes
\[
O\Bigg(\frac{n\|B_{\bullet F}\|_3^3\max_k \sigma_k^2}{\epsilon_3^2} + n^2\Bigg)
=O\Bigg(\frac{n\kappa^2 \|B_{\bullet F}\|_3^3 \min_k \sigma_k^2}{\epsilon_3^2} + n^2\Bigg)
=O\Bigg(\frac{\kappa^2 \|B_{\bullet F}\|_3^3 \|A\|_F^2}{\epsilon_3^2} + n^2\Bigg).
\]

\section{Estimating the error in proposition \ref{prop:state prepartion}}
\label{app4}

We only consider the case when all given data are real.
By measuring (\ref{alg: state prepartion}), if we get $|1\rangle$, then the post measurement state is
\[|\psi''\rangle=\frac{1}{\sqrt{Y}}\sum_{k=0}^{n-1}b_k\sin(f(k)t)|k\rangle,\]
where $Y=\sum_{k=0}^{n-1}|b_k\sin(f(k)t)|^2$. The desired state is
\[|\psi'\rangle=\frac{1}{\sqrt{Z}}\sum_{k=0}^{n-1}b_kf(k)|k\rangle,\]
where $Z=\sum_{k=0}^{n-1}|b_kf(k)|^2$. Then
\[\ba{lll} \vspace{.1cm}
\langle\psi''|\psi'\rangle &=& \ds\frac{1}{\sqrt{YZ}} \sum_{k=0}^{n-1}b_k^2f(k)\sin(f(k)t)
\geq \ds\frac{1}{\sqrt{YZ}} \sum_{k=0}^{n-1}b_k^2f(k)\Big(f(k)t-\frac{\epsilon_1^3}{6}\Big) \\ \vspace{.1cm}
&=& \ds\frac{1}{\sqrt{YZ}} \Big(Zt-\frac{\epsilon_1^3}{6}\sum_{k=0}^{n-1}b_k^2f(k)\Big)
\geq \ds\frac{1}{\sqrt{YZ}} \Big(Zt-\frac{\epsilon_1^3}{6}\sqrt{Z}\Big)
=\ds\frac{1}{\sqrt{Y}} \Big(\sqrt{Z}t-\frac{\epsilon_1^3}{6}\Big) \\ \vspace{.1cm}
&\geq& \ds\frac{1}{\sqrt{Z}t} \Big(\sqrt{Z}t-\frac{\epsilon_1^3}{6}\Big)
= 1-\ds\frac{\epsilon_1^3}{6\sqrt{Z}t}
\geq 1-\ds\frac{\epsilon_1^2\kappa(f)}{6},
\ea\]
where in the second step, we use the fact $\sin x\geq x-x^3/6$ and $|f(k)t|\leq \epsilon_1$.
In the fourth step, we applies Cauchy-Schwarz inequality to
$$\sum_{k=0}^{n-1}b_k^2f(k)\leq\sqrt{\sum_{k=0}^{n-1}b_k^2}\sqrt{\sum_{k=0}^{n-1}b_k^2f(k)^2}=\sqrt{Z}.$$
In the sixth step, we use the fact that $Y\leq Zt^2$.
And in the final step, we applies $|f(k)t|\geq \epsilon_0$ for all $k$
and $\epsilon_0=\kappa(f)^{-1}\epsilon_1$. Therefore, we have
\[\||\psi''\rangle-|\psi'\rangle\|_2=\sqrt{2(1-\langle\phi|\psi'\rangle)}\leq\sqrt{\frac{\kappa(f)}{3}}\epsilon_1=O(\epsilon),\]
if we choose $\epsilon_1=O(\epsilon/\sqrt{\kappa(f)})$.

\end{document}